# Thermal Accelerated Aging Methods for Magnet Wire: A Review

Lukas L. Korcak  
Institute of Technology Carlow,  
Kilkenny Road,  
Carlow, Ireland.

Darren F. Kavanagh  
Institute of Technology Carlow,  
Kilkenny Road,  
Carlow, Ireland  
darren.kavanagh@itcarlow.ie

*Abstract*— This paper focuses on accelerated aging methods for magnet wire. Reliability of electrical devices such as coils, motors, relays, solenoids and transformers is heavily dependent on the Electrical Insulation System (EIS). Accelerated aging methods are used to rapidly simulate the conditions in real life, which is typically years (20,000 hours) depending on the operating conditions. The purpose of accelerated aging is to bring lifetime of an EIS to hours, days or weeks. Shortening the lifetime of an EIS to such an extent, allows for the study of the insulation materials behavior as well as investigate ways to estimate the remaining useful life (RUL) for the purpose of predictive maintenance. Unexpected failures in operation processes, where redundancy is not present, can lead to high economical losses, machine downtime and often health and safety risks. Conditions, under which thermal aging methods are generally reported in the literature, typically neglect other factors, owing to the sheer complexity and interdependence of the multifaceted aging phenomena. This paper examines some existing thermal aging tests, which are currently used to obtain data for enamel degradation in order to try to better understand of how the thermal stresses degrade the EIS. Separation of these stresses, which the EIS operate under, can yield a better understanding of how each of the Thermal, the Electrical, the Ambient and the Mechanical (TEAM) stresses behave.

## I. INTRODUCTION

Accelerated aging methods for an EIS looks to rapidly simulate the conditions, whereby the enamel wire insulation degrades at a much faster rate than normal operating conditions in a real application [1]. During this time period various different properties of the test specimens are measured [2].

Empirical studies of accelerated aging with respect to temperature is well documented in the literature [3-8]. However, these methods and approaches have some shortcomings and limitations.

As aging is a multifaceted problem, these accelerated aging (AA) methods do not cover all of the operating conditions such as thermal cycling, electrical stresses, water ingress, humidity and chemical exposure, vibrations and mechanical stresses. In addition, it would appear that there is not any viable non-destructive measuring method or test, which would give a reliable estimate of EIS remaining life, as high voltage breakdown studies are typically utilised in industry. Also it needs to be pointed out, that if the *real-world* operating conditions change (e.g. humidity, temperature, vibration, pressure) with respect to the previous *laboratory* operating conditions, the lifetime duration can be dramatically shorten and hence reach catastrophic failure much earlier than anticipated. This paper reviews the various different methods and results for thermal accelerated aging from the literature and identifies some contradictions found within independent experimental findings.

## II. AGING METHODS

There are various standards and frameworks published by ASTM, NEMA, UL and IEEE for aging methods, measurement techniques and statistical analysis. These standards were established for assessing an EIS quality for industrial use in electric machines. Accordingly, different classes of wire as defined in Table 1 have different aging temperature limits.

As when the different stresses are combined in the AA, this does not allow for the impact of each individual stress to ascertained. In addition, the aging process within an EIS cannot be accurately represented by a simple equation due to the large of amount of variables contained in a process of the degradation. However, simplified equations, statistics and rules of thumb are often used in industry despite their inaccuracy as it is typically not feasible to factor in the specific aspects of the end application. It is clear that this is a highly complex and multifaceted area and one that is not well understood. Hence, more experimental studies and multi-physics modeling should be investigated to factor in the various aspects of electric machine design, coil and wire geometries, drive cycles and the various environmental conditions present.

## III. THERMAL CLASSES OF ENAMEL WIRE

Thermal endurance standard test method, permits comparison of the temperature/time characteristics for a film of insulated round enamelled wire for different thermal classes as defined in Table 1. An Arrhenius plot of lifetime vs temperature is typically used [2], [9]. The estimated endurance time for a specified temperature is usually 20,000 hours [9], [10].

Heat shock test measures the capability to resist cracking of the thin-film insulation on magnet wire after physical stressing and after a rapid temperature change. Capability of heat shock shall be at least 20 ºC higher than the wire insulation type class temperature [10]. Table 2 shows the heat shock parameters from ASTM standard D2307.

TABLE I. THERMAL CLASSES IN CELSIUS AND THE VARIOUS INSULATION TYPES [9], [10]

| Thermal class °C | Thermal classes | |
|---|---|---|
| | *Note* | *Insulation Types* |
| 90 | - | Paper |
| 105 | - | Paper, cotton |
| 105 | Solderable | Polyurethane |
| 105 | - | Nylon |
| 105 | - | Polyvinyl Acetal, Formvar |
| 130 | - | Epoxy |
| 130 | Solderable | Polyurethane |
| 155 | - | Polyester |
| 155 | Solderable | Polyurethane |
| 155 | - | Glass |
| 180 | - | Polyester |
| 180 | Solderable | Polyurethane |
| 180 | Solderable | Polyester-Imide |
| 200 | - | Polyester/amide imide, AI |
| 220 | - | Amide-Imide |
| 240 | - | Polyimide |

TABLE II. HEAT SHOCK PARAMETERS FOR THE ASTM STANDARD D2307, WIRE GAUGE IN AWG, % ELONGATION AND THE MANDREL DIAMETER [9], [10]

| Heat shock | | |
|---|---|---|
| *AWG* | *% Elongation* | *Mandrel Diameter* |
| 4-9 | 30 | n/a |
| 10-13 | 25 | 5d |
| 14-30 | 20 | 3d |
| 31-44 | 20 | 3d |

## IV. THERMAL TESTING

Temperature is increased on tested specimens above their rated operating temperature. Specimens are heated inside thermal ovens, alternatively heat could be created by passing through high electric currents with respect to the cross-sectional area of the wire, this is known as Joule heating [11]. These increased temperatures cause an accelerated degradation of the EIS.

The effects on the conductor material cannot be left out. Copper is the most common conductor material employed and its oxidation rate depends on conductor purity, temperature and oxygen permeability of varnish and enamel [3]. Temperatures above 200 ºC has a severe effect on chemical oxidation processes.

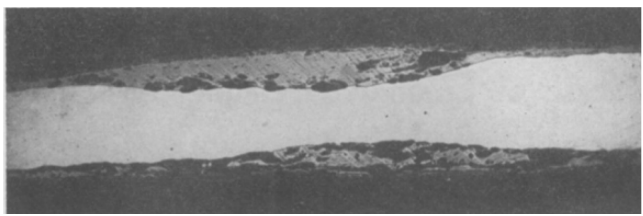

Fig. 1. Copper oxidation under enamel layer [3]

This reduces cross sectional area and increases the resistance of a conductor see Fig. 1. It also leads to an increase in the heat generation which in turn further accelerates the oxidation process [3]. The oxidation rate on the EIS depends on an enamel and temperatures involved. Experiments conducted by Petitgas et al. [4] presented on Fig. 2, show relative weight stability of an insulation at lower temperatures. Accordingly, the weight loss rate increases respectively with the temperature, however we can observe that once a critical point is reached, rapid degradation takes place which causes the pyrolysation of the EIS. Moreover, this will happen regardless, even if the EIS were to be placed within an inert atmosphere. However, this critical point for an inert atmosphere is higher [4].

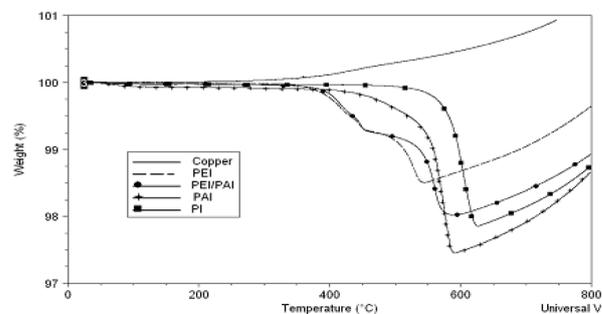

Fig. 2. Thermogravimetry (TG) curves of enameled wire and copper under air [4]

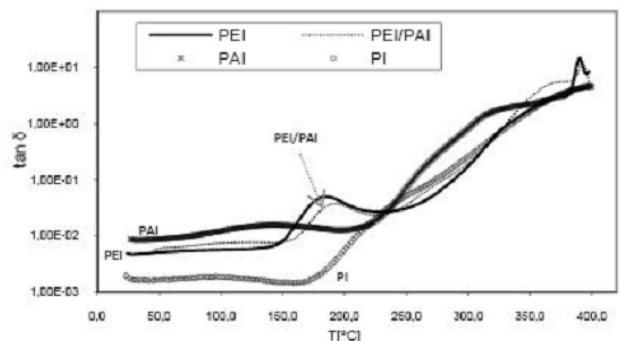

Fig. 3. Dielectric loss factor from room temperature to 400 C under $N_2$ atmosphere at 1 kHz [4]

Elevated temperatures may also have a post curing effect on enamels and cause a shifting of the dielectric relaxation phenomena within certain materials [12-14]. Dielectric relaxation spectroscopy results are shown on Fig. 3. Glass transition temperature within an enamel increases with aging time and temperature value respectively. This will have an effect on the mechanical properties of an enamel [4], changes to the temperature alone, results in changes of the material properties such as ductility, elongation and volume increase due to heat expansion.

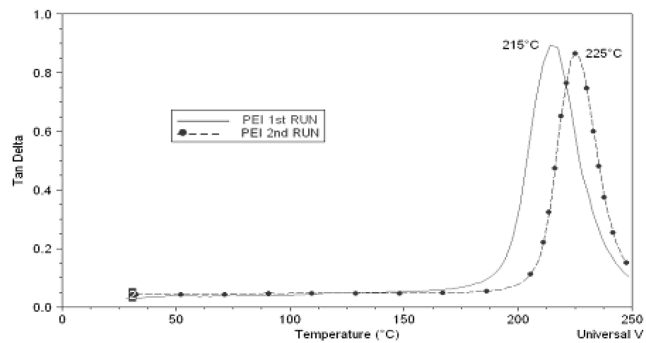

Fig. 4. Dynamic Mechanical Analysis presenting post curing change [4]

The degradation of EIS polymers during oxidation causes a shortening of the molecular chains during the cooling process, this causes a decrease in the dielectric properties [15].

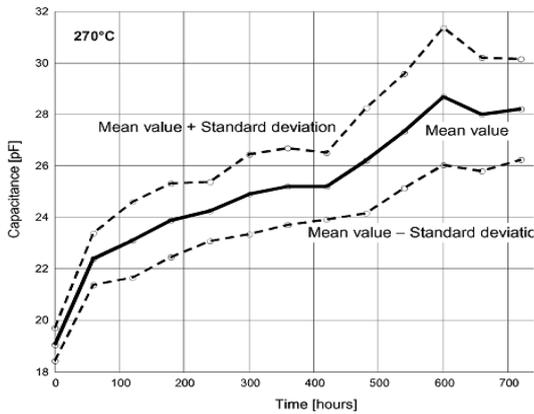
Fig. 5. Capacitance change during thermal aging of twisted pairs [5]

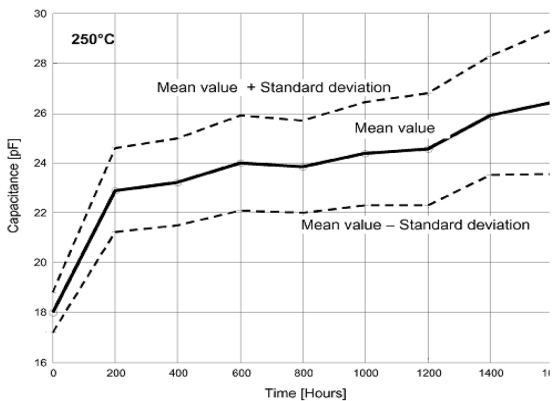
Fig. 6. Capacitance change during thermal aging of twisted pairs [5]

Sumislawska et al. [16] presents a nonlinear decrease of capacitance on tested specimens Fig. 7, however Savin et al. [17], [18] and Werynski et al. [5] showed, that insulation capacitance is increasing during the thermal aging process with 12 cycles (Fig. 5 and Fig. 6). Their work incorporated the IEC 60216-1, 2 and 3 standards in the experimental design.

The authors could not find much work that investigates the impact of thermal shocks on the aging process, with the exception being some studies conducted for the aircraft industry [6].

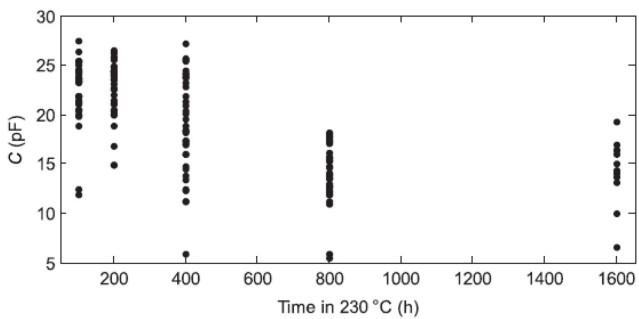
Fig. 7. Capacitance as function of time for insulation aged in 230 ᵒC [16]

## V. DISCUSSION

Standards were created for AA and testing, outlining procedures for obtaining required results of aged insulating materials. Accelerated aging techniques for all stresses are described in detail by these standards and are constantly updated, whenever better test methods are created [9]. However, not all researchers follow these guidelines and that may be a possible reason for some result contradictions occurring in research literature.

Enamel insulations are divided into different classes with respect to temperature, in which they can operate without premature failure. In this maximum allowed temperature, insulation should have a lifetime in the region 20,000 hours [9], [10]. Using different EIS materials has an impact on the cost of an electric machine. It is not necessary to use high temperature materials, in operating conditions, where temperatures are not as high. Similarly, this approach applies to the enamel grade (insulation thickness).

Accordingly, increasing the enamel grade will lead to an overall increase in the cost of the EIS, but also affect the machines 'fill factor'. This will have an effect on the power and torque density of the resulting machine. Hence original equipment manufacturers (OEMs) try to use wires with the least amount of enamel coating on them that is feasible for the temperature and voltage levels concerned for a given reliability.

The aging of the copper itself may have a substantial influence on EIS measurements during aging process. It was described in [3] where significant changes were found. A longitudinal cut of an enamelled wire after aging was shown in Fig. 1, indicating copper oxidation under enamel. It was stated, that temperatures over 200 ºC has a serious impact on copper conductors, especially when there are more impurities present in the copper [3]. It is likely that this copper oxide layer will have an impact on the measured capacitance and insulation resistance due to chemical changes present in the specimens. Reducing the cross-sectional area would lead to local hotspots and further more increase copper oxidation speed [3]. Over time this could significantly increase the temperature levels above EIS rated class and thus would cause the premature failure of the insulation.

Weight loss of EIS using thermogravimetry was conducted by Petitgas et al. which was presented in Fig. 2, shows the effects of elevated temperatures on different EIS materials. High temperatures cause the evaporation of enamels at certain temperatures and even pyrolysation to occur [4]. Dielectric loss of tested wire samples is apparent from Fig. 3. showing the increases in Tan Delta [4]. The theory would suggest that the aging of the polymeric enamels should result in an overall decrease of in the measured enamel capacitance. Capacitance decrease as a result of aging was presented in [16]. However as seen on Fig. 5 and Fig. 6 from [5], which follows the IEC 60851-5 standard procedure, capacitance was shown to increase in value during the aging process. This is very contradicting result which should be further verified. The difference could be as a result of different aging conditions.

Sumislawska et al. [16] stated, that delaminating insulation was observed at 800h of aging under temperature of 230 ºC. In Werynski et al. [5] experiment (Fig. 5 and Fig. 6), unusual capacitance change occurred between 600h and 800h, which was not explained. These results combined with copper oxidation described in [3], could lead to conclusion, that copper oxidation layer was formed and caused delaminating of enamel. Starting at 600h for over 200 ºC temperature of exposure. Since measurement of capacitance wasn't done for 600h in [16] (Fig. 7), but capacitance trend did not change significantly from 800h onwards, hence only assumptions for this event can be made. Closer inspection of this change should be made and establish whenever it has or has not impact on insulation aging. Delaminating of an insulation could have significant impact on the other TEAM stresses and promote their effects.

As a summary of this study, experiments closely following the established standards, should be made simultaneously with experiments designed by researchers, to ensure that a deviation for results was not biased by other factors.

## VI. Conclusion

This paper has presented an overview of results, for some accelerated aging of magnet wire found in the literature. The methods and approaches found in papers appear not to be able to accurately assess condition of an EIS. In addition, some contradictions of results could be found.

Much of the literature studied on thermal aging neglects the other aspects such as electrical, ambient and mechanical stresses acting on EIS. Also wire enamels experience measureable property changes during initial temperature increases which is caused by post curing. In addition, if the temperature is elevated too much, it will drastically shorten the lifetime, however the results are not meaningful, as other non-linear phenomena will occur such as increased evaporation and pyrolysation of the enamel. Copper oxidation also occurs between the copper substrate and the enamel coating during thermal AA, however it is unclear how this effects the real operational life of the EIS.

Different ways of assessing insulation health should be explored further and ideally encompass non-destructive methods. The challenge remains to design AA approaches that age the materials in such a way that closely mimics the real-world conditions. Furthermore, future work should explore unified methods for AA, so results and findings from the experiments do not contradict, unless special conditions were introduced into the AA process.